\documentclass[runningheads]{llncs}

\usepackage{graphicx}
\usepackage{hyperref}
\usepackage{color}
\usepackage{xcolor}
\begin{document}

\title{
Efficiently Distributed Federated Learning 
\thanks{This work receives EuroHPC-JU funding under grant no. 101034126, with support from the Horizon2020 programme (the European PILOT) and from the Spoke "FutureHPC \& BigData” of the ICSC – Centro Nazionale di Ricerca in "High-Performance Computing, Big Data, and Quantum Computing", funded by European Union – NextGenerationEU.}
}

\titlerunning{Efficiently Distributed Federated Learning}

\author{
Gianluca Mittone\inst{1}\orcidID{0000-0002-1887-6911} \and
Robert Birke\inst{1}\orcidID{0000-0003-1144-3707} \and
Marco Aldinucci\inst{1}\orcidID{0000-0001-8788-0829}
}

\authorrunning{G. Mittone et al.}

\institute{
Computer Science Department, University of Turin, Turin, Italy\\
\email{\{gianluca.mittone,robert.birke,marco.aldinucci\}@unito.it}\\
}

\begin{center}
The following paper is the accepted version of Springer copyrighted material
\\[12pt]
\textit{Gianluca Mittone, Robert Birke, \& Marco Aldinucci (2023). Efficiently Distributed Federated Learning. In Euro-Par 2023: Parallel Processing Workshops - Euro-Par 2023 International Workshops, Limassol, Cyprus, August 28 - September 1, 2023, Revised Selected Papers, Part II (pp. 321–326). Springer.}
\\[12pt]
presented at the EuroPar'23 conference in Limassol, Cyprus.
\\[12pt]
DOI: \href{https://doi.org/10.1007/978-3-031-48803-0_40}{https://doi.org/10.1007/978-3-031-48803-0\_40}
\end{center}

\maketitle

\begin{abstract}
Federated Learning (FL) is experiencing a substantial research interest, with many frameworks being developed to allow practitioners to build federations easily and quickly. 
Most of these efforts do not consider two main aspects that are key to Machine Learning (ML) software: customizability and performance. 
This research addresses these issues by implementing an open-source FL framework named FastFederatedLearning (FFL). 
FFL is implemented in C/C++, focusing on code performance, and allows the user to specify any communication graph between clients and servers involved in the federation, ensuring customizability. 
FFL is tested against Intel OpenFL, achieving consistent speedups over different computational platforms (x86-64, ARM-v8, RISC-V), ranging from 2.5x and 3.69x.
We aim to wrap FFL with a Python interface to ease its use and implement a middleware for different communication backends to be used.
We aim to build dynamic federations in which relations between clients and servers are not static, giving life to an environment where federations can be seen as long-time evolving structures and exploited as services.
\end{abstract}

\section{Introduction}
\subsubsection{Distributed Machine Learning}
\emph{Federated Learning} (FL)~\cite{mcmahan2017communication} is a \emph{Machine Learning} (ML) paradigm in which different entities collaborate to train a ML model without sharing their data. 
In \emph{Edge Inference} (EI), conversely, multiple entities participate in the inference over a dataset, each running the inference process independently with the local model. 
We refer to FL, EI, and other distributed approaches to ML as \emph{Distributed Machine Learning} (DML)~\cite{verbraeken2020survey} approaches. 
In this paper, we focus primarily on FL and EI, which are also \emph{federated approaches}, in which data is not shared among the clients.

\subsubsection{Limitation of current frameworks}
Each DML paradigm can be described as a \emph{communication graph}.
For example, FL implies a directed cyclic graph, while EI requires a directed acyclic one. 
The majority of current FL frameworks are strict in this sense: each one implements a small subset of communication graphs, like FedML~\cite{chaoyanghe2020fedml} and NVIDIA FLARE~\cite{roth2022nvidia}, and usually only one, like Intel OpenFL~\cite{reina2021openfl}, and FLOWER~\cite{beutel2020flower}; this means that the federation structure is hard-coded in the framework itself, and a peer-to-peer federation created by HP Swarm Learning~\cite{warnat2021swarm} cannot be replicated with Intel OpenFL, and vice-versa.
Furthermore, users cannot create \emph{customized communication graphs}, and not all frameworks allow \emph{dynamic nodes} to join and leave the federation at will.

Another limitation of current FL frameworks is that many support only \emph{Deep Neural Networks} (DNNs), with only a few supporting other ML models.
The main non-deep ML models currently supported are Gradient Boosting Decision Tree (GBDT), like FATE~\cite{liu2021fate}, and eXtreme Boosting (XGBoost) of Decision Trees, like NVIDIA FLARE.
Other \emph{traditional ML models} could be federated due to their properties and performance, but merging different models may require different algorithms and communication protocols, and current FL frameworks hard-code these aspects of the DML process.

Finally, current FL frameworks do not consider \emph{computational performance}.
While major ML libraries offer a Python interface that covers the complexity of the underlying C/C++ high-performance implementation, current FL frameworks are written in plain Python.
While it is true that the underlying ML library does the bulk of the computation, it is also necessary to take into account the distributed structure of a DML system and the subsequent waiting times and delays introduced by poor distributed programming.
Efficient communication structures and data serialisation libraries will remove unnecessary waiting times and improve computational power usage, improving training times.

\section{FastFederatedLearning}
To address the above mentioned issues, we propose \emph{FastFederatedLearning}~\footnote{\url{https://github.com/alpha-unito/FastFederatedLearning}} (FFL)~\cite{mittone2023ffl}, an open-source, high-performance, C/C++ based FL framework.
FFL exploits the FastFlow~\cite{aldinucci2017fastflow} parallel programming framework for handling both single-node simulations and real-world distributed runtimes, the Cereal~\cite{grant2013cereal} library for data serialisation, and the PyTorch C++ interface~\cite{NEURIPS2019_9015} for training.
In its current implementation, FFL already achieves faster execution times when compared to commercial FL software like Intel OpenFL, but it does not yet fully address all identified main issues of FL frameworks.

\subsubsection{Custom communication graph}
\label{sec:custom}
While already able to express various communication graphs, FFL still lacks a few features.
Imitating most ML software on the market, we aim to hide the high-performance C/C++ code of FFL under a user-friendly Python-based hood.
Using a Keras-like syntax, the user can specify a complete, full-scale federation with a small set of keywords and methods, just like is commonly done with DNNs in frameworks such as PyTorch and TensorFlow~\cite{tensorflow2015-whitepaper}.
This high-level description will be translated into a formal representation based on the RISC-pb$^2$l~\cite{aldinucci2014design} formal language, which can be mapped to FastFlow code in a one-to-one fashion.
This formal representation can be simplified and optimized by an ad-hoc symbolic reasoner before being converted into code to improve the system's expected performance as much as possible.
Such a \emph{middleware} would also allow different backends to be used. 
By constructing an ad-hoc interpreter of RISC-pb$^2$l, users would have the possibility to use different parallel and distributed programming frameworks outside FastFlow.
This feature opens the possibility of experimenting with different communication libraries. 
Commercial FL frameworks almost only rely on the gRPC+protobuf combo to handle communication. 
FastFlow instead supports MPI and TCP, and many other communication protocols, such as MQTT, could be envisioned.

%
Many studies~\cite{briggs2020federated,sattler2020clustered,ghosh2020efficient} currently target the FL learning performance by applying clustering techniques to the clients involved in the process.
In this case, the clients are often selected based on their data distribution, computational power, or even model architecture and are sorted out between different servers to obtain better aggregated models.
Many of these approaches either require organizing clients in different clusters according to the state of the federation or having a hierarchical structure to handle the process better.
To support these use cases, we plan to expand FFL by introducing a third entity in the structure other than server and clients, namely a \emph{proxy node}.
This new kind of node will be in charge of intermediating between server and clients, taking care of doing operations as partial aggregation, update encryption, supervising the federation process, storing the history of updates to back up in case of poisoning and other attacks, or just forwarding updates to the central server.
Furthermore, as presented in the next paragraph, we will experiment with dynamic communication graphs.
This way, we can manage the clusters at runtime, moving clients dynamically across different servers and proxies according to the algorithmic and federation needs.

\subsubsection{Dynamic communication graph}
\label{sec:dynamic}
FFL still cannot handle dynamic nodes due to FastFlow, which creates a static communication graph.
We want to experiment with \emph{dynamic communication graphs} in the DML scenario, allowing the movement of nodes across different servers.
Such an approach would have a practical reflection in the IoT scenario, in which each sensor would then have the possibility to connect to the nearest server, or also in the automotive one, where moving across geographically distant zones could require changing the server connection.
This dynamicity will open the door to experimenting with new types of FL algorithms, such as the clustering one mentioned in the previous paragraph, but also with long-living federations, in which different federations live over longer periods, merging, swapping nodes and being created and destroyed continuously according to the needs of the participating institutions.

%
Furthermore, having a dynamic communication graph will open the possibility of exploiting \emph{FL-as-a-Service} (FLaaS)~\cite{kourtellis2020flaas}.
This new take on FL consists of offering the membership to the federation to a client as a possible service to the client itself.
This approach can be exploited, for example, by large organisations federated together to share their FL model with smaller ones, thus making it exploit their high-quality ML model for the period agreed.
After the closing of the service, the small organisation will be excluded from the federation, thus having no longer access to the updated shared model.
Large dataset owners organised in federations can exploit this business model, and small organizations can exploit the shared knowledge without needing access to large datasets and huge computational power.
Furthermore, organisations can not just share their model but also offer to train the client's model on their local dataset, thus offering data access and computational power as a service in a federated fashion. 

\subsubsection{Model Agnostic Federated Learning}
Most current FL frameworks support only DNN models due to the very nature of those structures: being representable as tensors, they are easy to serialise, send over the network, and aggregate straightforwardly (e.g., calculating their mean).
Also, DNNs usually offer high learning performance but require large datasets and high computational power to train efficiently.
Traditional ML models, on the other side, provide a better complexity-to-learning performance trade-off and can thus be used where DNNs are not applicable.
Recent research proposes an algorithm to allow the federation of any traditional ML model: AdaBoost.F~\cite{polato2022boosting}.
AdaBoost.F is a federated extension of AdaBoost and, thus, is a \emph{model-agnostic} algorithm: it can be applied using any ML model as a weak learner.
Such a feature requires the use of advanced serialisation tools and also to consider the computation's communication costs more.
Also, being different from the standard Federated Averaging (FedAvg) algorithm, AdaBoost.F requires a different communication protocol, which is usually hard-coded in current FL frameworks.

We implemented a Python-based, open-source version of AdaBoost.F into the Intel OpenFL FL framework: we called this software \emph{OpenFL-extended}~\footnote{\url{https://github.com/alpha-unito/OpenFL-extended}} (OpenFL-x).
To our knowledge, OpenFL-x is the first FL framework to be model-agnostic, supporting both DNNs and traditional ML models.
We have succeeded in optimising OpenFL-x by fine-tuning its distributed structure (e.g., refining the communication buffer sizes, the serialisation tool, and the sleeps present in the code) and also its internal structure (e.g., the internal tensor memorisation database) achieving a 5.5x speedup over the base software; however, it is still not able to scale over a large number of computational nodes efficiently due to its design architecture.
We aim to implement the AdaBoost.F algorithm into the FFL framework, thus further empowering the possibilities offered by it, combining at the same time new and innovative ways of running FL with the additional benefit of a high-performance C/C++ implementation at scale.

\subsubsection{High-performance Federated Learning}
As can be seen, our leitmotiv is ``\emph{high-performance C/C++ implementation}''.
We believe that current Python-based FL frameworks introduce inefficiencies in the FL process, slowing down the computation and convergence time of the models.
All current widely exploited ML frameworks use the same strategy: masking the underlying complexity of a C/C++ implementation with a user-friendly Python interface, thus obtaining an excellent trade-off between the performance and usability of their software.
From our perspective, FL frameworks are not different: they should not introduce more delays than the strictly necessary ones needed for communications.
We thus aim to provide a Python-based, Keras-style interface to ease the use of FFL, making it more appealing and user-friendly, as discussed earlier. 

Another contribution that we want to integrate into FFL are \emph{asynchronous communications} between the different components of the federation.
This approach is starting to get studied~\cite{chen2020asynchronous,lu2020privacy,wu2020safa}, but still lacks a mature FL framework allowing the user to choose whether or not to use asynchronous communications.
The motivation behind this is clear: while asynchronous communication makes distributed computation faster, they are harder to manage and introduces ML-related issues, such as merging new and old updates from different sources.
We aim to implement asynchronous FL algorithm(s) already validated by the literature into FFL and to test them in a real-world, distributed scenario.

We tested FFL on a heterogeneous cluster, successfully running FL training on a federation comprising x86-64, ARM-V8, and RISC-V devices: we assessed a 2.5x to 3.69x speedup with respect to OpenFL.
This \emph{real-world testing phase} is not trivial since most FL-related papers do not validate their methods in the wild but instead simulate them on small, homogeneous clusters, abstracting from issues like unstable internet connections, slow or delayed devices, high communication costs between long distances, differences in computation/communication capabilities, and different computational architectures. 
We aim to test FFL with different strategies available in the literature for exploiting such real-world federations at their best, e.g. not waiting for stragglers or slow devices.

\section{Conclusions}
We presented the current FL frameworks scenario, emphasising their limits and proposing possible solutions.
We identify as significant research areas in this field the lack of dynamicity and customizability of the federation communication graph, the need to support traditional ML models other than DNNs, and the opportunities given by high-performance, asynchronous implementation of such frameworks.
We propose FFL as an in-development research software for experimenting with these concepts.

%
%
\bibliographystyle{Styles/splncs04}
\bibliography{Bibliography/biblio}
\end{document}